\documentclass[prl,twocolumn,showpacs]{revtex4}
\usepackage{graphicx,epsf}
\usepackage{bm, bbm}      % bold math

\newcommand{\beq}{\begin{equation}}
\newcommand{\beqn}{\begin{eqnarray}}
\newcommand{\eeq}{\end{equation}}
\newcommand{\eeqn}{\end{eqnarray}}

\begin{document}

\title{Bridge between Abelian and Non-Abelian Fractional Quantum Hall States}
\author{N. Regnault$^1$, M. O. Goerbig$^2$, and Th. Jolicoeur$^3$}

\affiliation{
$^1$Laboratoire Pierre Aigrain, D\'epartement de Physique, ENS, CNRS, 24 Rue Lhomond, F-75005 Paris, France\\
$^2$Laboratoire de Physique des Solides, CNRS UMR 8502, Univ. Paris-Sud, F-91405 Orsay cedex, France\\
$^3$Laboratoire de Physique Th\'eorique et Mod\`eles Statistiques, Univ. Paris-Sud, F-91405 Orsay cedex, France}

\begin{abstract}

We propose a scheme to construct the most prominent Abelian and non-Abelian
fractional quantum Hall states from $K$-component Halperin wave functions. In order to account
for a one-component quantum Hall system, these SU($K$) colors are distributed over all 
particles by an appropriate symmetrization. Numerical calculations corroborate
the picture that $K$-component Halperin wave functions may be a common basis for
both Abelian and non-Abelian trial wave functions in the study of one-component 
quantum Hall systems.

\end{abstract}
\maketitle

Trial wave functions play a central role in the study of the fractional quantum Hall effect (FQHE). Thus,
the FQHE at filling factors $\nu\equiv n_{el}/n_B=1/3$, $1/5$, ..., in terms of the electronic $n_{el}$
and the flux $n_{B}=eB/h$ densities, respectively, has been explained with the help of Laughlin's 
wave function \cite{laughlin}. The states of the principle FQHE series $\nu=p/(2sp\pm 1)$ have 
found a compelling interpretation in terms of composite-fermion (CF) wave functions \cite{Jain89}. Other
proposals include non-Abelian states, which may be described with the help of a Pfaffian wave function
at half-filling \cite{MR} or a parafermionic generalization of it \cite{Read99}, at $\nu=K/(K+2)$, in terms of
an integer $K$. All of these states may be viewed as sophisticated generalizations of Laughlin's wave
function to describe incompressible quantum liquids.

In addition to these one-component wave functions, which treat fermions of only one type 
(spinless fermions), a two-component generalization has been proposed by Halperin to account 
for the spin degree of freedom \cite{halperin}. A similar situation arises in bilayer quantum Hall
systems where the bilayer index may be mimicked by an isospin, and the FQHE may 
equally be described in terms of two-component Halperin wave functions \cite{perspectives}. 
Multicomponent quantum Hall systems have recently attracted increasing interest because of
the discovery of a particular quantum Hall effect in graphene, two-dimensional graphite \cite{QHEGraphene}.
Due to its fourfold spin-valley degeneracy, graphene in a strong magnetic field may indeed
be viewed as a four-component quantum Hall system \cite{GrapheneRev}.

Motivated by graphene, as well as
bilayer quantum Hall systems with non-polarized spin, two of us have recently generalized
Halperin's wave function to the case with $K$ different components (``colors'') \cite{HalperinSU4}.
Here, we show that these SU($K$) Halperin wave functions may also play an important role in
the understanding of the original one-component quantum Hall system, once symmetrized with respect 
to their color degree of freedom. This symmetrization of the SU($K$) Halperin wave functions 
allows one to obtain the non-Abelian Read-Rezayi (RR) states at $\nu=K/(K+2)$ as well as 
its straight-forward generalization to $\nu=K/(nK+2)$ \cite{Read99}. We corroborate this statement within
exact-diagonalization (ED) studies on the sphere. Similarly, one may
obtain the states at $\nu=K/(2sK+1)$ of the CF series if one multiplies the SU($K$)
Halperin states by a product of $K(K-1)/2$ permanents. Within Monte-Carlo calculations, 
we show that these states have a reasonably large overlap, for different accessible system sizes,
with CF wave functions \cite{Jain89}.

We consider the $K$-component Halperin wave function
\beq\label{wave}
\Psi^{(K)}_{[m;n]} = \prod_{k<l}(z_k-z_l)^n\prod_{i=1}^K\prod_{k_i<l_i}^{N/K}
\left(z_{k_i}^{(i)}-z_{l_i}^{(i)}\right)^{m-n},
\eeq
where $z_{k_i}^{(i)}$ is the position of the $k_i$-th particle with color $(i)$ in
the complex plane. Here, $N$ is the total number of particles, and the number of
particles per color is, thus, $N^{(i)}=N/K$.
The first Jastrow term is SU($K$)-symmetric and, thus, does not distinguish 
between different particle colors. Explicitly it may be written as
$
\prod_{k<l}(z_k-z_l)^n \equiv \prod_{i<j}\prod_{k_i,k_j}\left(z_{k_i}^{(i)}
-z_{k_j}^{(j)}\right)^n\prod_{i}\prod_{k_i<l_i}\left(z_{k_i}^{(i)}-z_{l_i}^{(i)}\right)^n.
$
Eq. (\ref{wave}), where we have omitted an overall Gaussian factor, represents a particular
form of SU($K$) Halperin wavefunctions \cite{HalperinSU4} where all intra-color correlations
are the same, with an exponent $m$, and all inter-color correlations are described by
the exponent $n$. In this form, it describes FQHE states at a total filling factor \cite{Degail08}
\beq\label{filling}
\nu=\frac{K}{nK+(m-n)}\qquad {\rm with} \qquad \delta_{[m;n]} =m ,
\eeq
where $\delta_{[m;n]}$ denotes the shift in the sphere geometry used in our ED
calculations.

One notices that this state corresponds to a state of high symmetry because the different 
color groups are equally occupied, i.e. the generalized polarizations of the SU($K$)
spin are all zero. These polarizations are associated with
the elements of the Cartan algebra, the set of the $K-1$ generators of SU($K$) that mutually commute.
The state (\ref{wave}) is, therefore, invariant with respect to color permutations
[Weyl group of SU($K$)]. We implement this symmetry in the ED studies and use 
appropriate Haldane pseudopotentials \cite{haldane} to generate the SU($K$) Halperin state, as
described in Ref. \cite{HalperinSU4}.

{\sl -- Non-Abelian states.}
A compelling approach to construct the bosonic non-Abelian RR states
at $\nu=K/2$ has been provided by Cappelli {\sl et al} \cite{Cappelli01}. They may 
indeed be obtained from the 
generalized Halperin wave function $\Psi_{[2;0]}^{(K)}$,
$
\Psi^{(K)}_{RR}=\mathcal{S}\;\Psi^{(K)}_{[2;0]},
$
where the $\mathcal{S}$ symbol stands for the symmetrization over all possibilities to associate $K$ colors 
to all $N$ particles. A straightforward generalization to fermions 
is obtained with the help of the $\Psi_{[3;1]}^{(K)}$ state,
\begin{eqnarray}
\Psi^{(K)}_{RR}&=&\mathcal{S}'\;\Psi^{(K)}_{[3;1]}\ ,
\label{cappellifermions}
\end{eqnarray}
where $\mathcal{S}'$ has a similar meaning as $\mathcal{S}$, except that it removes those cases, which 
are equivalent up to global color permutation in order to avoid accidental cancelation due to the inherent 
fermionic nature of the $[3;1]$ state. Notice that $\mathcal{S}'$ may also be used for a bosonic state
if one changes its normalization. 
Very recently, the states (\ref{cappellifermions}) and their degeneracy 
have been studied on the torus \cite{Ardonne08}.

The symmetrization procedure may be applied directly
to any $N$-body SU($K$) Fock state. It consists of discarding any component of this
state, which has at least one orbital with an occupancy greater than one, and of summing all components
with the same orbital occupancy regardless of their color pattern. Furthermore, if we consider linear
combinations of $N$-body states which are invariant under the discrete symmetries of the Weyl group, only
the maximally symmetric ones survive this symmetrization procedure.

In order to test the above construction, we have checked by ED that one reproduces,
up to machine precision,
the RR states at $\nu=K/(K+2)$ from their Halperin counterparts for $K=2$ 
(for $N=4$, $6,...12$ particles) at $\nu=1/2$ (Pfaffian state), 
$K=3$ ($N=6$, 9, and 12) at $\nu=3/5$, and $K=4$ ($N=4$ and $N=8$) at $\nu=2/3$. 
We emphasize the large Hilbert space dimension 
due to the internal degree of freedom; e.g., for $K=3$ and $N=12$, the dimension of the largest subspace is 
$212\,121\,434$ if one accounts for all discrete Weyl symmetries of $SU(3)$. 
Thus, this method to generate RR states is more involved as compared to using an appropriate
$(K+1)$-body  interaction (see e.g. \cite{Simon07}) or the ``squeezing'' technique \cite{Bernevig07}.

Beyond these ground-state properties, Capelli {\sl et al.} conjectured that the 
RR quasihole states may be obtained from the corresponding Halperin wave functions \cite{Cappelli01}. 
We have checked this hypothesis by ED for $K=2$, 3, and 4 in the case of the 
$[3;1]$ state. 
If one adds one flux quantum to the RR state, one generates $K$ quasiholes of a particular
degeneracy \cite{Ardonne02}, which may be sorted with respect to the eigenvalues of the 
angular momentum operators $L^2$ and $L_z$. In contrast to these states, the addition of one
flux quantum to the SU($K$) Halperin wave function $[3;1]$ generates a larger set of degenerate 
quasiholes. However, we find that the degeneracy counting with respect 
to $L^2$ and $L_z$ is the same if we limit ourselves to the maximally symmetric sector of the Weyl group.

Notice that if both the Halperin and the RR states had the same degeneracy for any number, larger than one, 
of added flux quanta, this would preclude non-Abelian quasiparticle statistics because of the inherent
Abelian nature of the excitations of the Halperin state. This is, however, not the case -- the degeneracies
are different for two added flux quanta. The symmetrization, thus, removes all unwanted states 
in the SU($K$) case, and one obtains the correct degeneracy of the RR quasihole states, in agreement with
their non-Abelian character. For a complete understanding of quasihole excitations in the framework
of Halperin wave functions, additional conformal-field theoretical (CFT) studies would be required, which are
beyond the scope of the present paper.

\begin{figure}
\centering
\includegraphics[width=5.5cm,angle=0]{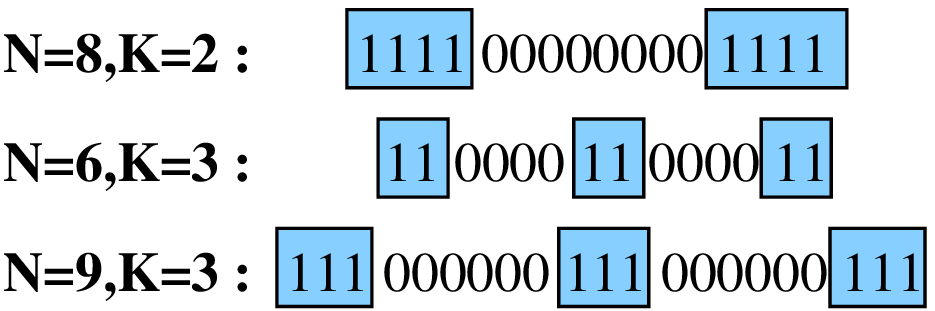}
\caption{\footnotesize{Thin-torus limit of $\mathcal{S}'\Psi_{[1;3]}^{(K)}$ for different
values of $K$ and $N$. The particles phase-separate into $K$ repelling droplets, as expected for
$m<n$. In general, these highest weights and the ``squeezing'' technique \cite{Bernevig07} do 
neither lead to a unique quantum state nor a state identical to that obtained by color symmetrization.
}}
\label{fig01}
\end{figure}

Beyond the $K$-color $[3;1]$ states, which reproduce to great numerical accuracy the RR states
at $\nu=K/(K+2)$, one may investigate a generic $[m;n]$ state. Consistency of the 
symmetrization procedure requires $m$ and $n$ to be both odd (even) in a fermionic (bosonic) 
one-component state. The $1/m$ Laughlin state is reproduced for $m=n$. For $m<n$, the $[m;n]$ state is 
unstable in the generalized 2D plasma picture, and the (classical) particles of
different colors undergo a phase-separate \cite{Degail08}. However, the quantum state 
$\mathcal{S}'\Psi^{(K)}_{[m;n]}$ is, in view of the analyticity condition for the
lowest Landau level (LL),
a valid candidate and one may look at the thin torus limit i.e. the highest weight 
as defined in \cite{Bernevig07}. 
In this limit, we clearly see the analogue of the plasma instability
(Fig. \ref{fig01}): the $N$ particles tend to form $K$ mutually repelling clusters of $N/K$ particles. 
Moreover, even if $\mathcal{S}'\Psi^{(K)}_{[1;3]}$ has the same filling factor and shift as the 
non-Abelian states with reverse flux attachment \cite{Jolicoeur07}, the overlap 
is low ($0.659$ for $K=3$ and $N=6$). 
Therefore, such states may not be related to homogeneous FQHE states.

\begin{figure}
\centering
\includegraphics[width=7.5cm,angle=0]{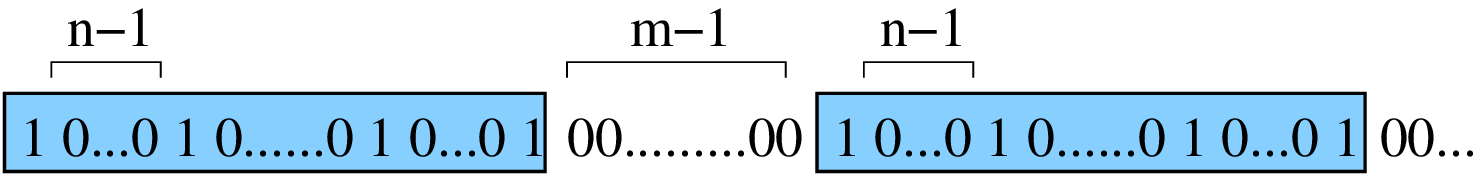}
\caption{\footnotesize{Thin-torus limit of $\mathcal{S}'\Psi_{[m;n]}^{(K)}$. The shaded boxes 
indicate clusters of $K$ particles. The spacing between the clusters is determined by the exponent $m$,
whereas that within each cluster by $n$, i.e. the inter-color correlations. $n=0$ 
(for bosons) corresponds to particles on the same site within each cluster.
}}
\label{fig02}
\end{figure}

The $m>n$ case is stable with respect to phase separation \cite{Degail08} and is, thus, a promising
candidate for a homogeneous state.
If we choose $m-n=2$, which also accounts for the above-mentioned case $m=3$, $n=1$, we 
obtain generalized RR states \cite{Read99} at filling factors $\nu=K/(Kn+2)$. 
One may, therefore, conjecture that for every pair $[m;n]$ ($m>n$) with the same parity, the 
trial wave function 
$\mathcal{S}'\Psi^{(K)}_{[m;n]}$ yields a series of states at filling factors 
$\nu=K/(Kn+r)$, the excitations of which may obey non-Abelian statistics ($n$ even or odd for 
bosons or fermions, respectively). Here, we have defined $r=m-n$.
Fig. \ref{fig02} shows the generic highest weight limit of the states $\mathcal{S}'\Psi^{(K)}_{[m;n]}$
which we have obtained by ED for different values of $m$, $n$, $K$, and $N$. We have 
checked numerically that the application of the ``squeezing'' technique \cite{Bernevig07} 
applied to this highest weight for $m=5$, $n=1$
($K=2$) yields the same quantum state as $\mathcal{S}'\Psi^{(2)}_{[5;1]}$. 
More generally, squeezing spans a subspace which $\mathcal{S}'\Psi^{(K)}_{[5;1]}$ belongs to.
The pattern consists of $N/K$ equally spaced clusters with $K$ particles per cluster. Whereas the 
separation between the particles within each cluster is $n-1$ (the bosonic $n=0$ 
represents a cluster where all $K$ particles are on the same site), the inter-cluster separation
is governed by the intra-color correlation exponent $m$. Before symmetrization, the particles per 
cluster may be viewed as $K$ particles of different color. This must be contrasted to the 
unstable case $m<n$, discussed above, where $N/K$ particles of the same color clusterize into 
$K$ repelling droplets. This indicates a certain duality, where the unstable Halperin states may
be viewed as (stable) Halperin states of $N/K$ pseudo-colors with $K$ ``particles'' per color.

In order to obtain insight into the statistical properties of the quasihole excitations of the
$[5;1]$ states, we have investigated their degeneracies when adding one or more flux quanta (for
$K=2$ and 3). Above a certain number of added flux quanta, the degeneracies of the states
$\mathcal{S}'\Psi^{(K)}_{[5;1]}$ are indeed different from $\Psi^{(K)}_{[5;1]}$ restricted to
the maximally symmetric sector. Without being a strong proof, these results indicate that
the quasihole excitations may be non-Abelian, as for the RR states.

Very recently, Bernevig and Haldane \cite{Bernevig08} have proposed a similar series for bosonic states
at $\nu=K/r$, using Jack polynomials. This series matches ours for $n=0$, and both types of states have
the same shift on the sphere ($\delta=r$). We have checked numerically that both states 
have the same thin-torus limit. 
However, in our case, $r$ must be even and we recover only half of their states. 

{\sl -- States of the CF series.}
When $m$ and $n$ have different parity, the symmetrization procedure fails in generating fermionic or 
bosonic states. This problem may be fixed by
multiplying the $\Psi_{[m;n]}^{(K)}$ state with the correct object which restores the right statistics 
for the final one-component wave function. To illustrate this approach, we first consider a modified 
$K=2$ $[3;2]$ Halperin wave function at $\nu=2/5$,
\beq
\tilde{\Psi}^{(2)}_{[3,2]}
={\rm perm}\left[M^{(1,2)}\right] \Psi^{(2)}_{[3,2]}\left(z^{(1)}_1,...,z^{(1)}_N,z^{(2)}_1,...,z^{(2)}_N\right)
\label{cf332su2}\ ,
\eeq
in terms of the $N\times N$ matrix 
\beq
M^{(i,j)}_{k_i,k_j} = \left[z^{(i)}_{k_i} - z^{(j)}_{k_j}\right]^{-1} ,
\label{MSU2}
\eeq
$i,j=1,2$ here,
and ${\rm perm}\left[M^{(1,2)}\right]=\sum_{\{\sigma\}}\prod_{k=1}^N\,M^{(1,2)}_{k,\sigma(k)}$ is 
the permanent of  $M^{(1,2)}$, where the sum is over all $\sigma$ permutations of $N$ elements. $\mathcal{S}'\tilde{\Psi}^{(2)}_{[3;2]}$ corresponds to a fermionic two-component
state. The symmetrized wave function (\ref{cf332su2}), as a candidate for the FQHE at $\nu=2/5$ 
has been studied by Yoshioka {\sl et al.}, and one obtains a large overlap with the ED 
ground state for a Coulomb interaction \cite{Yoshioka88}. This wave function, also called 
``Gaffnian'', has recently been studied within CFT and may support non-Abelian  
quasi-particle excitations \cite{Simon07b}.

To generalize this construction to higher values of $K$, we propose the following wave function
\beq
\tilde{\Psi}^{(K)}_{[3,2]} = \prod_{i<j}{\rm perm}\left[M^{(i,j)}\right]\Psi^{(K)}_{[3;2]},
\label{cf332suK}
\eeq
in terms of a product of $K(K-1)/2$ permanents of the matrix (\ref{MSU2}).
These wave functions may describe states at filling factors $\nu=K/(2K+1)$, which corresponds to the
principle series of two-flux CF \cite{Jain89}. 
Notice further that both $\mathcal{S}'\tilde{\Psi}_{[3;2]}^{(K)}$ and Jain's CF wave
function are found at the same shift $\delta=K+2$ on the sphere.
The utility of the wave function (\ref{cf332suK})
as an alternative to Jain's CF construction was already pointed out by Morf in 2000 \cite{Montreux}. 
Moreover, the wave function (\ref{cf332suK}) is similar, but not equivalent to a generalization of 
the Gaffnian \cite{Simon07b}. The function (\ref{cf332suK}) may be related to $r=K+1$ and 
the highest weights as described in Fig. \ref{fig02} (see \cite{Bernevig08} for $K=2$).
However, it has been shown that the CFTs associated with Gaffnian-type functions 
are non-unitary and may, thus, yield to critical states with vanishing gap
\cite{Simon07b,Read07}.

By Monte-Carlo integration, we have studied quantitatively the overlap between 
$\mathcal{S}'\tilde{\Psi}_{[3;2]}^{(K)}$ and the CF wave function $\Psi_{CF}^{(K)}$
for $K$ filled pseudo-LLs 
\cite{Jain97}, for $K=2$ ($\nu=2/5$) and $K=3$ ($\nu=3/7$) and various values of $N$. The results
are displayed in Tab. \ref{overlapcf}. The overlaps for all studied system sizes are above 95\%. The strong statement
that both approaches are equivalent, as indicated by the large overlap of the different trial wave 
functions, would, however, require a detailed comparison of the quasiparticle excitations and their
statistical properties, which is beyond the scope of this paper.
Nevertheless, we stress that the generic $2s$ flux CF also fits within the 
multicomponent scheme if we choose $n=2s$ and $m-n=1$. 

%%%%%%%%%%% vieux %%%%%%%%%%%%%%%%%%%
\iffalse
One may conjecture that the reverse flux 
attachment may be obtain by $m-n=-1$ and by inverting all individual elements of  $M^{(i,j)}$ since both the filling factor and the shift match their CF 
analogues. However, the overlap with the CF wave function 
at $\nu=2/3$ drops 
from $0.901(3)$ for $N=6$ down to $0.097(2)$ for $N=12$. 
This corroborates the argument to discard unstable Halperin wave functions.
\fi
%%%%%%%%%%%%%%%%%%%%%%%%%%%%%%%%%

\begin{table}
\begin{tabular}{c|c|c|c|c|c}
$N$ & 6 & 8 & 10 & 12 & 14\\
\hline
${\cal O}_{\nu =2/5}$ & 0.991 (3) &  0.982 (4) & 0.977 (5) & 0.972 (5) & 0.968 (6)\\
\hline\hline
$N$ & 6 & 9 & 12 & 15\\
\hline
${\cal O}_{\nu =3/7}$ & 0.993 (3) & 0.979 (4) &  0.963 (5) & 0.954 (7)\\
\end{tabular}
\caption{\footnotesize Overlap between the CF wave functions $\Psi_{CF}^{(K)}$ with $K$ 
filled pseudo-LLs and $\mathcal{S}'\tilde{\Psi}_{[3;2]}^{(K)}$
for $K=2$ (at $\nu=2/5$, upper row) and $K=3$ (at $\nu=3/7$, lower row). 
${\cal O}_{\nu}=|\langle \Psi_{CF}^{(K)}| \mathcal{S}'\tilde{\Psi}_{[3;2]}^{(K)}\rangle|$ 
is computed using Monte-Carlo integration 
with $10^6$ iterations. The numerical error on the last digit is indicated in parenthesis.}\label{overlapcf}
\end{table}

In order to obtain a better understanding of the relation between CF trial wave functions and 
$\mathcal{S}'\tilde{\Psi}_{[2s+1;2s]}^{(K)}$, we rewrite the SU($K$) Halperin wave function as 
a product of two others, $ \Psi_{[2s+1;2s]}^{(K)}=\Psi_{[2s;2s]}^{(K)}\Psi_{[1;0]}^{(K)}$.
The SU($K$)-symmetric one, $\Psi_{[2s;2s]}^{(K)}$, mimics the attachment of $2s$ flux quanta per
particle of any color and is, thus, common to both approaches, and $\Psi_{[1;0]}^{(K)}$
is a product of $K$ Slater determinants for each color, i.e. the particles of each color 
form a liquid at an effective color filling $\nu_i^*=1$ with no correlations between 
different colors. Beyond this similarity with CF wave functions, we emphasize the following 
difference: in the CF approach, one describes particles of only one color which fill $K$
pseudo-LLs. This generates non-analytic components in the wave function, which must thus
be projected to the lowest LL \cite{Jain89}. In our case, the trial wave function is
naturally in the lowest LL, but the price to pay
is the introduction of $K$ artificial colors, accompanied by the symmetrization 
procedure. Furthermore, in contrast to Jain's wave functions and the above non-Abelian states,
one needs to introduce the permanent in order to describe particles with well-defined statistical
properties.

We have, furthermore, investigated states with $m-n=-1$, the filling factor and shift on the
sphere of which match those of their CF analogues with reversed flux attachment. 
However, the overlap with the CF wave function at $\nu=2/3$ drops 
from $0.901(3)$ for $N=6$ down to $0.097(2)$ for $N=12$. 
This corroborates the argument to discard unstable Halperin wave functions, as for the
description of non-Abelian states.

In conclusion, we have shown that $K$-component Halperin wave functions 
$\Psi_{[m;n]}^{(K)}$, with $m>n$ yield a common basis for the description 
of one-component FQHE states. The color symmetrization of states with 
$m$ and $n$ of the same parity ($m,n$ odd for fermions and $m,n$ even for 
bosons) yields the generalized non-Abelian RR states for $m-n=2$.
If $m$ and $n$ have different parity and if the Halperin wave function 
$\Psi_{[m;n]}^{(K)}$ is multiplied by a product of $K(K-1)/2$ permanents with respect to the
different colors, the symmetrization yields wave functions of high overlap with
Jain's CF states at $\nu=K/(nK+1)$, for even $n$ and $m-n=1$. The colors correspond to
pseudo-LLs in the CF picture, but the wave functions $\Psi_{[m;n]}^{(K)}$
are fully in the lowest LL and need, thus, not be projected. A natural question arises
whether a generalization to odd values of $r=m-n\neq 1$ may describe further CF-type states
at $\nu=K/(nK+r)$ \cite{Slingerland08}.

We acknowledge fruitful discussions with A. Bernevig, R. Santachiara, and J. Slingerland.
We thank R. Morf for communicating us his talk at the 2001 EPS conference in Montreux and S.
Simon for pointing out the relation of Eqs. (\ref{cf332su2}) and (\ref{cf332suK}) with the generalized Gaffnian.
This work was funded
by the Agence Nationale de la Recherche under Grant No. ANR-07-JCJC-0003-01.


\begin{thebibliography}{99}

\bibitem{laughlin}R.\ B.\ Laughlin, Phys.\ Rev.\ Lett.\ {\bf 50}, 1395 (1983).

\bibitem{Jain89} J.K. Jain, Phys. Rev. Lett. {\bf 63}, 199 (1989).

\bibitem{MR}G.\ Moore and N.\ Read, Nucl.\ Phys.\ B\ {\bf 360}, 362 (1991).

\bibitem{Read99}N. Read and E. Rezayi, Phys. Rev. B {\bf 59}, 8084 (1999).

\bibitem{halperin}B. I. Halperin, Helv. Phys. Acta {\bf 56}, 75 (1983).

\bibitem{perspectives}For a review, see
{\sl Perspectives in Quantum Hall Effects}, edited by S.\ Das\ Sarma and
A.\ Pinczuk (John Wiley, New York, 1997).

\bibitem{QHEGraphene}K. S. Novoselov, A. K. Geim, S. V. Morosov, D. Jiang,
M. I. Katsnelson, I. V. Grigorieva, S. V. Dubonos, and A. A. Firsov,
Nature {\bf 438}, 197 (2005);
Y. Zhang, Y.-W. Tan, H. L. Stormer, and P. Kim,
{\sl ibid.} 201.

\bibitem{GrapheneRev}A. H. Castro Neto, F. Guinea, N. M. R. Peres, K. S. Novoselov, and
A. K. Geim, arXiv:0709.1163.

\bibitem{HalperinSU4}M.\ O.\ Goerbig, and N.\ Regnault, Phys.\ Rev.\ B\ {\bf 75}, 241405(R) (2007).

\bibitem{Degail08}R. de Gail, N. Regnault, and M. O. Goerbig, Phys.\ Rev.\ B\
{\bf 77}, 165310 (2008).

\bibitem{haldane}F. D. M. Haldane, Phys. Rev. Lett. {\bf 51}, 605 (1983).

\bibitem{Cappelli01} A. Cappelli, L. S. Georgiev and I. T. Todorov, Nucl. Phys. B {\bf 599}, 499 (2001).

\bibitem{Ardonne08}E. Ardonne, E. J. Bergholtz, J. Kailasvuori, and E. Wikberg,
J. Stat. Mech. P04016 (2008).

\bibitem{Simon07} S.H. Simon, E.H. Rezayi, N.R. Cooper, Phys. Rev. B {\bf 75}, 195306 (2007).

\bibitem{Bernevig07} B. A. Bernevig and F. D. M. Haldane, arXiv:0707.3637.

\bibitem{Ardonne02}E. Ardonne, J. Phys. A {\bf 35} 447 (2002); N. Read, Phys. Rev. B {\bf 73}, 245334 (2006).

\bibitem{Jolicoeur07} Th. Jolicoeur, Phys. Rev. Lett. {\bf 99}, 036805 (2007).

\bibitem{Bernevig08} B. A. Bernevig and F. D. M. Haldane, arXiv:0803.2882.

\bibitem{Yoshioka88}D. Yoshioka, A. H. MacDonald, and S. M. Girvin, Phys. Rev. B {\bf 38}, 3636 (1988).

\bibitem{Simon07b}S. H. Simon, E. H. Rezayi, N. R. Cooper, and I. Berdnikov,
Phys. Rev. B {\bf 75}, 075317 (2007).


\bibitem{Montreux}R. Morf, talk at the European Physical Society Conference of the
Condensed Matter Division (EPS-CMD18), Montreux 2000.

\bibitem{Read07}N. Read, arXiv:0711.0543.

\bibitem{Jain97} J.K. Jain and R.K. Kamilla,
Int. J. Mod. Phys. B {\bf 11}, 2621 (1997);
Phys. Rev. B {\bf 55}, R4895 (1997).

\bibitem{Slingerland08} see also E. Ardonne and J. Slingerland, {\it in preparation}.

\end{thebibliography}
\end{document}